\documentclass[fleqn,usenatbib]{mnras}

\usepackage[T1]{fontenc}
\usepackage{ae,aecompl}

\usepackage{graphicx}	
\usepackage{txfonts}                  

\newcommand{\HI}{H{\,\small I}}
\newcommand{\FRI}{FR{\small I}}
\newcommand{\FRII}{FR{\small II}}
\newcommand{\msun}{M_{\odot}}
\newcommand*\rad{~rad\,m$^{-2}$}

\DeclareRobustCommand{\ion}[2]{%
\relax\ifmmode
\ifx\testbx\f@series
{\mathbf{#1\,\mathsc{#2}}}\else
{\mathrm{#1\,\mathsc{#2}}}\fi
\else\textup{#1\,{\mdseries\textsc{#2}}}%
\fi}

\newcommand{\OIII}{[\ion{O}{iii}]~$\lambda$4959, 5007} 
\newcommand{\OII}{[\ion{O}{ii}]~$\lambda$3727} 

\title[Faraday rotation study of NGC\,612]{Faraday rotation study of NGC\,612 (PKS\,0131-36): a hybrid radio source and its magnetised circumgalactic environment}
\author[Banfield et al.]{J.~K. Banfield$^{1,2}$, S.~P.~O'Sullivan$^{3}$, M.~H. Wieringa${^4}$ and B.~H.~C.~Emonts$^{5}$\\
$^{1}$ Research School of Astronomy and Astrophysics, Australian National University, Canberra, ACT 2611, Australia\\
$^{2}$ ARC Centre of Excellence for All-Sky Astrophysics (CAASTRO)\\
$^{3}$ Hamburger Sternwarte, Universit\"at Hamburg, Gojenbergsweg 112, Hamburg 21029, Germany\\
$^{4}$ CSIRO Astronomy and Space Science, P.O. Box 76, Epping, NSW 1710, Australia\\
$^{5}$ National Radio Astronomy Observatory, 520 Edgemont Road, Charlottesville, VA 22903
}
\begin{document}

\date{Accepted 2018 November 11. Received 2018 October 22; in original form 2017 November 13.}

\pagerange{\pageref{firstpage}--\pageref{lastpage}} \pubyear{2018}

\maketitle

\label{firstpage}

\begin{abstract} 
We present a polarization and Faraday rotation study of the hybrid morphology radio galaxy NGC\,612 (PKS\,0131$-$36), using Australian Telescope Compact Array observations from 1 to 3 GHz. 
In general, the results are consistent with an external Faraday screen close to the radio source. In the eastern \FRII\ lobe, the RM of the hotspot increases in magnitude towards the leading edge, as well as changing sign (compared to the rest of the lobe). The Faraday depolarization is also $\sim$3 times larger at the hotspot than elsewhere. A plausible explanation for this is significant compression of ambient magnetised gas by the bow shock produced by the advancing hotspot. 
The western \FRI\ lobe also exhibits some evidence of interaction with local magnetised gas, as a transverse band of high RM coincides with a distinct bend in the lobe. 
Previous observations of NGC~612 revealed an \HI\ bridge of tidal debris along the direction of the eastern lobe towards the gas-rich companion NGC~619. We find no clear evidence that ionised gas associated with this bridge is either mixing with or lies in the foreground of the radio source. This is consistent with the absence of \HI\ absorption against the hotspot, and indicates that the tidal debris must lie mostly behind the eastern lobe.
\end{abstract}


\section{Introduction}
The nature of the interaction of radio-loud active galactic nuclei (AGN) with their environment is a key process in how both galaxies and radio-loud AGN evolve. The interaction between the radio lobes, formed by relativistic jets emanating from the supermassive black hole, with the interstellar medium (ISM), the intercluster medium (ICM), and the intergalactic medium (IGM), effect how gas is moved around to enhance or inhibit star formation. 

Radio-loud AGN have been found to suppress star-formation and affect how the supermassive black hole and host galaxy co-evolve \citep[i.e.][]{Croton2006} while also providing mechanical heating to suppress the ICM cooling \citep[i.e.][]{McNamara2012}. Simulations by \citet{Mukherjee2016} of radio jets passing through a multi-phase ISM show that the low-power radio jets ($P_{\rm jet} \le 10^{43}\,$ergs$^{-1}$) will impact the host galaxy over a larger volume than their high-power counterparts. \citet{Morganti2013} suggest that interactions between the radio-loud AGN jets with the ISM are most important in the young stages of the activity while at larger distances ($10 - 100\,$kpc) the radio-loud AGN plays a `maintenance' role.

On kpc scales, the impact of supersonic jets onto the ISM/IGM can create cocoons of (shocked) gas and synchrotron emission with hotspot morphologies (\FRII\ radio galaxies), while entrainment of gas by slow-jets can lead to bright radio sources with edge-darkened lobes (\FRI\ radio galaxies). X-ray observations have shown radio emission and X-ray emission are anti-correlated in both galaxy groups and clusters \citep[i.e.][]{Carilli1994,Hardcastle2007b}. \citet{Croston2007, Croston2011} have also shown that shock heating from the radio-loud jets can affect environments in both massive galaxies and gas-rich mergers. 
This interplay between fuelling the AGN and the expansion of the radio-loud AGN into the ISM, IGM, and ICM can shape how galaxies evolve.  

Studying the gas and magnetic field properties in and around radio lobes and jets provides key constraints on the details of the physical interaction between the radio jets and lobes with their environment. 
\citet{Saikia2003} suggest a link between the radio jets and the surrounding ISM through the differences of the degrees of polarization asymmetry between compact and extended radio sources. Radio continuum observations of the isolated radio galaxy B2 0755+37 by \citet{Guidetti2012} show anisotropic fluctuations of the rotation measure (RM) distribution across the two radio lobes consistent with cavities of thermal gas evacuated by the radio lobes.

The intrinsic magnetic field structure of radio lobes is typically circumferential (i.e.~parallel to the total intensity contours) at the edges of the lobes \citep{leahy1986}. The jet magnetic field is typically longitudinal in powerful \FRII\ type jets \citep[][]{saikiasalter1988}, while in \FRI\ the inner part of the jet often has a longitudinal field before transitioning to a transverse field in the outer jet \citep{laingbridle2014}. High resolution studies of hotspots show a diverse range of structures that elude simple classification schemes, with similarly complicated behaviours in the associated polarization fraction and orientation \citep[e.g.][]{black1992,leahy1997}. However, for those sources with an unambiguous `primary' hotspot, the magnetic field is often approximately transverse to the observed (or inferred) jet direction, while a `secondary', more diffuse, hotspot can often have a field transverse to a line connecting the hotspots \citep{hardcastle1997}. This suggests the presence of significant field compression caused by, and aligned with, the shock front \citep[e.g.][]{miller1985}. 

The nearby powerful radio galaxy NGC\,612 shows an apparent alignment between the radio-loud AGN and an \HI\,bridge connected to a gas-rich companion \citep{Emonts2008}. Motivated by this, we observed NGC\,612 in full polarization. Our goal is to examine the interaction between the radio-loud AGN and its gaseous environment, particularly at the hotspot. NGC\,612 is a low redshift radio-loud AGN that allows for a detailed study of the environmental impact of radio-loud jets and lobes, which may be of additional use in interpreting the observed properties of high redshift systems.

NGC\,612 is a typical S0 galaxy at $z = 0.0297$ \citep{Westerlund1966,Goss1980,Veron2001} hosting a radio-loud AGN PKS 0131-36 \citep{Ekers1978}. PKS 0131-36 has been classified as a HYbrid MOrphology Radio Source \citep[HYMORS; ][]{Gopal2000}. The eastern radio lobe contains a hotspot near the furthest edge of the radio lobe typical of Fanaroff-Riley type II (\FRII) radio sources \citep{Fanaroff1974}.  The western lobe shows no indication of a hotspot and is more diffuse, resembling FR type I (\FRI) radio sources.  \citet{Ekers1978} observed PKS 0131-36 at 1415\,MHz with the Fleurs Synthesis Telescope and found that the monochromatic power across the whole radio extent of 400\,kpc is $P_{\rm 1.4\,GHz} = 8.4\times 10^{23}\,$W Hz$^{-1}$ ster$^{-1}$ with a luminosity density of $L_{\rm 1.4\,GHz}=1.6\times 10^{31}\,$W Hz$^{-1}$. \citet{Ekers1978} combined observations from the JVLA at 5\,GHz with observations at 5\,GHz and 8.1\,GHz by \citet{Schilizzi1976} to reveal a flat radio spectrum of the core that is compact at a limit of $< 0.3\arcsec$ centered on the host galaxy NGC\,612.

Observations by \citet{Westerlund1966} of NGC\,612 reveal a dust disk and noted the similarity to Centaurus A. \citet{Ekers1969} and \citet{Ekers1978} show that the radio axis is almost perpendicular to the dust disk at 78\,degrees. Emission line measurements of \OII, \OIII, and H$\alpha$ by \citet{Goss1980} uncovered a rotation curve out to approximately 40\,kpc at a maximum rotational velocity of 340\,km s$^{-1}$ with the rotation axis being perpendicular to the dust disk. Further investigation by \citet{Holt2007} indicate the presence of a young stellar population of 0.04 to 0.1\,Gyr throughout the stellar disk. Observations with {\it SUZAKU} from 0.5 -- 60\,keV by \citet{Eguchi2011} suggest NGC\,612 is a Compton-thick AGN with $N_{\rm H} \approx 10^{24}\,$cm$^{-2}$ similar to a Seyfert Type 2 galaxy.  The opening angle of NGC\,612 was estimated by \citet{Eguchi2011} to be 60 -- 70\,degrees.

The combination of observations of X-ray, radio, and optical in the literature show evidence of a possibly interacting galaxy group. \citet{Westerlund1966} determined there are 9 galaxies within $32\arcmin$ of NGC\,612. \HI\,\,observations by \citet{Emonts2008} reveal a faint \HI\,\,bridge extending 400\,kpc to nearby galaxy NGC 619 with no detectable optical counterpart related to the \HI\,\,bridge or tails. The \HI\,\,mass distribution is mainly found in disks around NGC 619 ($M_{\rm HI} = 8.9 \times 10^{9}\msun$) and NGC\,612 ($M_{\rm HI} = 1.8 \times 10^9\msun$), with $M_{\rm HI} = 2.9 \times 10^8\msun$ located in the \HI\,\,bridge connecting the two galaxies \citep[][see their Fig.~5]{Emonts2008}. X-Ray observations by \citet{Tashiro2000} detect diffuse emission at 0.7 -- 3\,keV extending away from NGC\,612 towards the eastern radio lobe and no emission near the western lobe. The hard X-Ray emission from 3 -- 10\,keV show a highly obscured active nucleus in NGC612 \citep{Tashiro2000}. 

In this paper, we present new broadband radio polarization and Faraday rotation observations of PKS 0131-36. We present our study as follows.  Sections \ref{sec:obs} and \ref{sec:imgs} provide details about the observations, calibration, and imaging of the radio data. Our analysis of these data are shown in section \ref{sec:analysis} with Section \ref{sec:discussion} providing the discussion. Throughout this paper we adopt a $\Lambda$CDM cosmology of $\Omega_M=0.3$, $\Omega_{\Lambda} = 0.7$ with a Hubble constant of $H_{\rm 0} = 70\,$km s$^{-1}$ Mpc$^{-1}$.  At the redshift of NGC\,612, $z=0.0297$, the luminosity distance is $D_L=130.1\,$Mpc and the angular size distance is $D_A=122.7\,$Mpc giving a scale of 0.595\,kpc arcsec$^{-1}$ \citep{Wright2006}.  We define the radio spectral index as $S_{\nu} \propto \nu^{\alpha}$.

\section{Observations and Data Processing}\label{sec:obs}
\subsection{Synthesis Observations}
The Australian Telescope Compact Array \citep[ATCA; ][]{Frater1992} with the Compact Array Broadband Backend \citep[CABB; ][]{Wilson2011} was used to produce a Stokes $I$ and linear polarization mosaic of NGC\,612 across the 16\,cm CABB band. We used the first intermediate frequency (IF) band for continuum observations with $2049\times1\,$MHz channels centered on the 16\,cm CABB band. 

Our observations were taken on 2012 October 25 -- 26 for 13 hours in the 750B array configuration under project ID C2728. The shortest baseline is 61\,m and the longest baseline is 4500\,m.  We used antenna 6 in our calibration.  However, we removed antenna 6 for our analysis leaving the longest baseline at 765\,m. The resulting angular resolution of our observations ranges from $92.86\arcsec \times 31.90\arcsec$ (position angle = -10.6\,deg) at 1.1\,GHz to $26.37\arcsec \times 14.82\arcsec$ (position angle = 0.8\,deg) at 3\,GHz. The primary beam full-width half-power (FWHP) ranges from $42\arcmin$ at 1.1\,GHz to $15\arcmin$ at 3.1\,GHz. A two pointing mosaic was made with one pointing centered on the East lobe of NGC\,612 and a second pointing centered on the West lobe. 
This was done becase the bright outer edges of the radio galaxy ($\sim$13$'$ angular size) extend close to the FWHP at the highest frequencies. The mosaic also helped minimise the effect of instrumental polarization across the extent of the source. At the time of our observations the instrumental polarization response was unknown across the 16\,cm CABB band both in frequency and distance from the pointing center. 

The primary flux calibrator, PKS~1934$-$638 \citep[14.95\,Jy at 1.380\,GHz; ][]{Reynolds1994}, was observed twice during the observations.  We observed B0048$-$427 as the phase calibrator for two minutes once every hour.  

\begin{figure}
\includegraphics[width=7.cm]{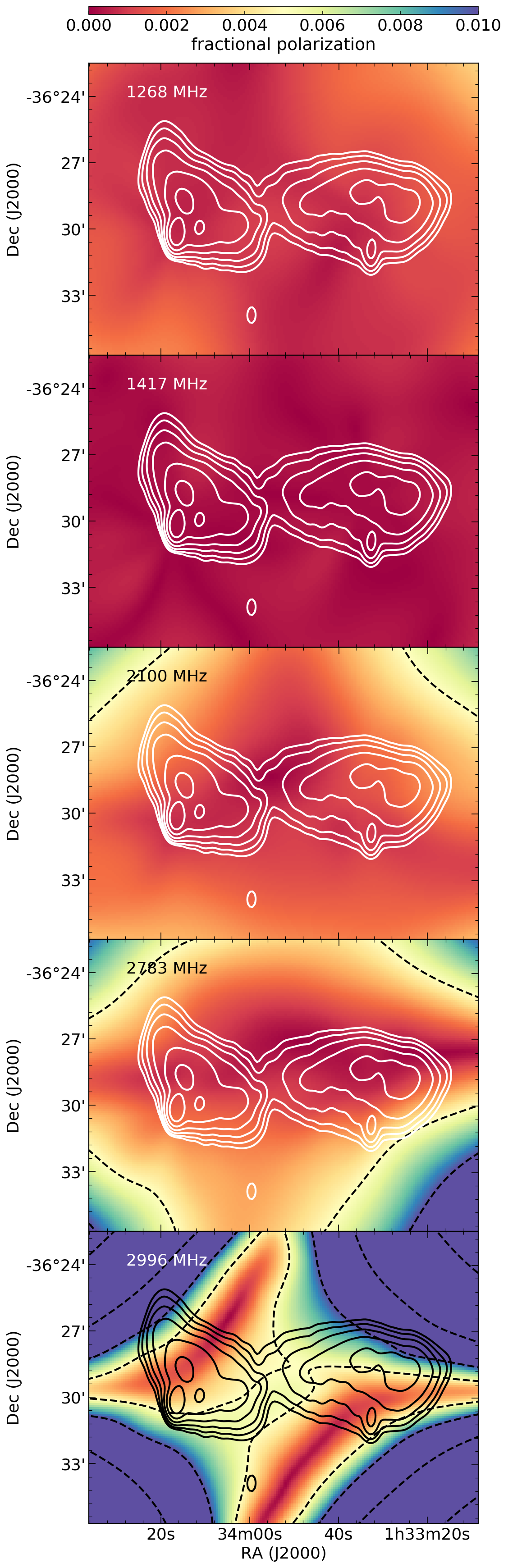}
\caption{The instrumental polarization response (colorscale) of the ATCA across the 16\,cm CABB band for our observations as described in Section \ref{sec:offpol}. The dashed contours mark the 0.5, 1, 2, and 5 per cent instrumental polarization.  
The Stokes $I$ data of NGC\,612 are shown with the solid contours, starting at $4\sigma_{\rm I}$ and increasing by a factor of 2. 
A color version of this figure is available online.}\label{fig:offpol}
\end{figure}

\subsection{Calibration}
Calibration and editing of the $uv$-data was completed in {\tt MIRIAD}\footnote{\href{http://www.atnf.csiro.au/computing/software/ miriad/}{http://www.atnf.csiro.au/computing/software/miriad/}} version 1.5 \citep{Sault1995}. The data were loaded into {\tt MIRIAD} using the task {\tt atlod} (v.1.52) and flagging of known radio frequency interference (RFI) was applied using the option {\tt birdie}.  The bandpass was calibrated with PKS~1934$-$638 using {\tt mfcal} (v.1.17). The frequency dependent gain and polarization leakage was solved for using the task {\tt gpcal} (v.1.22) in 8 bins of $256\,$MHz (using the parameter {\tt nfbin=8}). The bandpass and leakage solutions were then copied to the secondary calibrator B0048$-$427 using {\tt gpcopy} (v.1.12). The secondary was used to solve for the complex antenna gains as a function of time and for the polarization of the secondary, again in 8 frequency bins, using {\tt gpcal} \citep[see][]{Schnitzeler2013}. We then used task {\tt gpboot} to perform absolute flux calibration by scaling the secondary gains to those determined on the primary calibrator and used task {\tt mfboot} to correct any remaining error in the spectral slope. We expect the resulting flux density scale to be accurate to better than 5\% across the frequency band. Automated flagging of the calibrators was completed using {\tt pgflag} (v.1.30), the automatic flagging routine developed by \citet{Offringa2010} and incorporated into the {\tt MIRIAD} software package.  Manual flagging of the calibrators was also completed using {\tt uvflag} (v.1.7).
Once the calibration was complete, we copied the solutions to NGC\,612.  We then flagged NGC\,612 using {\tt pgflag} and manually flagged with {\tt uvflag}.  After flagging, the useable data were between $1.2 - 3.0\,$GHz.

\vspace{-0.4cm}
\subsection{Instrumental Polarization}\label{sec:offpol}
The main goal of our observations is to examine the rotation measure and possible interaction of the radio lobes of NGC\,612 with the environment.  In order to accurately test our hypothesis, we require measurements of the instrumental polarization response across the 16\,cm CABB band. We present our analysis in Figure \ref{fig:offpol} for five different frequencies (1268, 1417, 2100, 2783, and 2996\,MHz) with respect to our observations of NGC\,612. The instrumental polarization maps were created using the off-axis observations of PKS 1934-638 by J.~Stevens from 2011\footnote{http://www.narrabri.atnf.csiro.au/people/ste616/beamshapes/beamshape\_16cm.html }. We synthesised observations using the {\tt MIRIAD} task {\tt offpol} (v.1.8) with the new beam data assuming 12 hour observations with an hour angle range of $-6\,$h to $+6\,$h for each mosaic pointing for each of the five different frequencies in Stokes $I$, $Q$, and $U$.  The resulting images were mosaicked together using {\tt linmos} (v.1.32) and combined together to evaluate the fractional polarization at each frequency.

We found that the instrumental polarization response increases with frequency across the 16\,cm CABB band.  At 1268\,MHz, the fractional polarization is below 0.004 across the primary beam.  At the higher end of the band, the instrumental fractional polarization increases to 0.05 well within the primary beam.  In Figure \ref{fig:offpol} we overlay the Stokes $I$ flux density contours of NGC\,612 starting at the $4\sigma_{\rm I}$ level and increasing by a factor of 2.   At the low end of the band, there is less than 0.3 per cent instrumental polarization across NGC\,612.  In the middle of the band, the response increases to 0.5 per cent near the radio lobe edges of NGC\,612.  
At the top end of the band (2996\,MHz), the instrumental polarization does not exceed 2 per cent across NGC\,612 with the highest polarization leakage near the radio lobe edges. These maps are used in Section~\ref{sec:kmap} to provide an additional systematic uncertainty when modelling the fractional polarization across the 16~cm band. 

\begin{figure*}
\includegraphics[width=\textwidth]{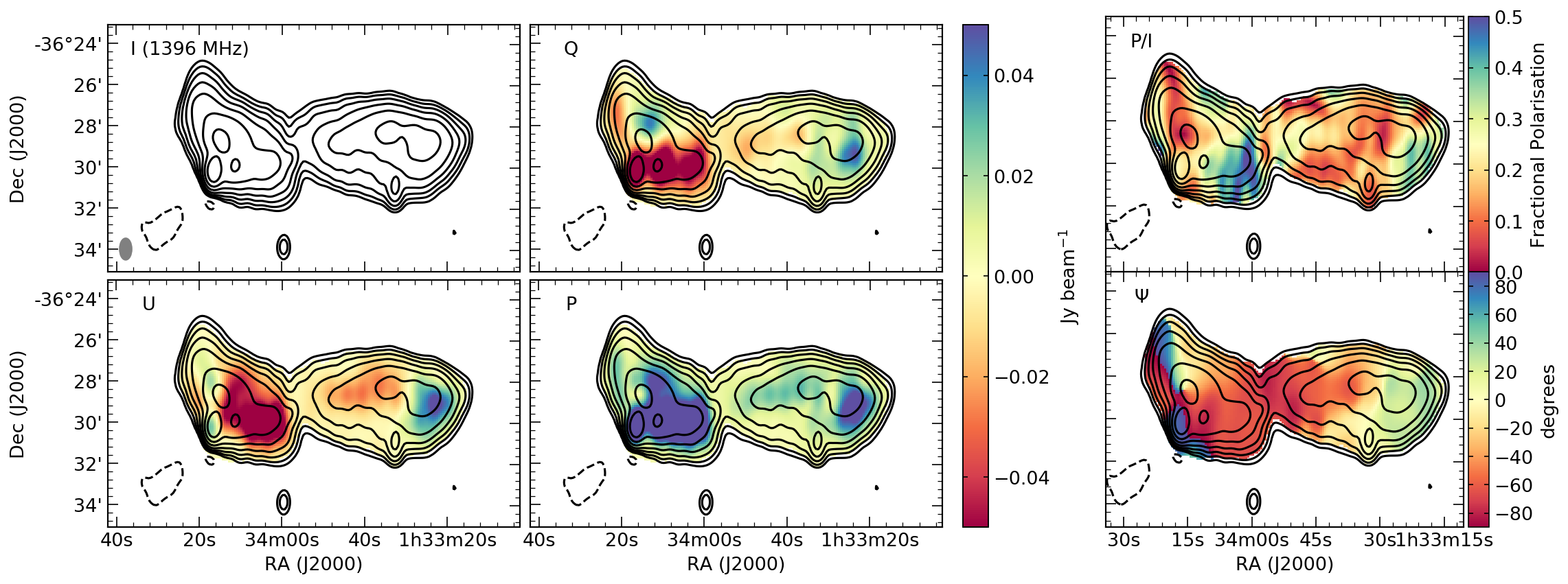}
\caption{The full Stokes ($I$, $Q$, $U$, $P$) images with polarization angle ($\Psi$) and fractional polarization ($P/I$) images at 1396\,MHz. The black contours are the Stokes $I$ radio emission from the full band continuum image with dashed contours at $-2\sigma_{\rm I}$ and solid contours at $2\sigma_{\rm I}$ and then increasing by factors of 2. The grey oval in the lower left of the Stokes $I$ image represents the beam. A color version of this figure is available online.}\label{fig:polimage}
\end{figure*}

\section{Images}\label{sec:imgs}
\subsection{Total Intensity}
Large fractional bandwidth poses three problems with wide-band wide-field imaging as noted by \citet{Condon2012}: (1) the field of view; (2) the point spread function; and (3) the flux densities of sources can vary significantly with frequency.  The fractional bandwidth of the $1.1-3.1\,$GHz CABB band is 95\,per cent and if we image the full band at once we would encounter the above imaging issues. We also encountered an additional issue with respect to our target source NGC\,612.  The radio lobes of NGC\,612 are well resolved and show complex structure at all spatial resolutions.  As a result, we took extra steps in imaging both in both continuum and linear polarization to decrease calibration and imaging artefacts. 

We used only antenna 6 baselines for the first rounds of phase self calibration, suppressing the resolved structure of the radio lobes and leaving the compact regions. We then excluded the antenna 6 data from further processing. For the second stage of imaging we used all 10 remaining baselines, but excluded the shortest two baselines in the self calibration. We carefully masked the emission regions to avoid sidelobes in the model. The Stokes $I$ continuum image was imaged in $3\times683\,$MHz bands, a compromise between limiting the fractional bandwidth to avoid deconvolution errors and retaining enough data for good imaging. 
The flux density variations of the source were handled with MFS imaging and deconvolution using {\tt MIRIAD} tasks {\tt invert} (v.1.22, with options {\tt mfs and sdb})  and {\tt mfclean} (v.1.11) to model the flux and spectral index in each of the three 683 MHz bands. 
The three frequency bands were convolved to the lowest resolution and combined in the image plane using {\tt linmos}. The final beam size is $58.8\arcsec \times 31.0\arcsec$ at $-0.75\,$degrees.  
The resulting Stokes $I$ noise level of $\sigma_{\rm I} = 1.59\,$mJy beam$^{-1}$, is $\sim$10 times larger than the theoretical thermal noise level, indicating that the image dynamic range is limited by calibration and imaging errors and not thermal noise. 

\subsection{Polarization}\label{sec:polarization}
We split the band into $16 \times 128\,$MHz bands to image polarization across the source.  Each 128\,MHz band was imaged separately and then convolved to the lowest resolution of $78\arcsec \times 35\arcsec$ at $-1\,$degree. Adopting the same notation and procedure as \citet{Sokoloff1998}, \citet{Farnsworth2011} and \citet{OSul2012} we define the complex linear polarization as:
\begin{equation}
P = Q + iU = pIe^{2i\Psi} = pIe^{2i(\Psi_0+{\rm RM}\lambda^2)} \,\,\, ,
\end{equation}
where $\Psi$ is the observed polarization angle and $I$, $Q$, and $U$ are the measured Stokes parameters.  The fractional polarization is defined as:
\begin{equation}
p = \sqrt{q^2 + u^2} \,\,\, ,
\end{equation}
where $q = Q/I$ and $u=U/I$. The linear polarization angle is 
\begin{equation}
\Psi = \frac{1}{2}{\rm arctan}\frac{u}{q} \,\,\, .
\end{equation}
The Faraday rotation measure (RM) rotates the linear polarization angle from its intrinsic value ($\Psi_0)$ as a function of wavelength-squared ($\lambda^2$).
The amount of Faraday rotation depends on the properties of magnetised plasma along the line of sight, as ${\rm RM} \sim 0.81\int^0_L n_e\,B_{||}\,dl$ with $n_e$ being the free electron number density (in cm$^{-3}$), $B_{||}$ the line-of-sight magnetic field strength (in $\mu$G) and $L$ the total path length (in parsecs). 

Figure~\ref{fig:polimage} shows the full Stokes ($I$, $Q$, $U$, $P$) images with polarization angle ($\Psi$) and fractional polarization ($P/I$) images at 1396\,MHz. The black contours are the Stokes $I$ radio emission from the full band continuum image starting at $4\sigma_{\rm I}$ and increasing by a factor of 2. 
Imaging using multi-frequency synthesis to better fill the $uv$-plane and obtaining the largest range in RM (by imaging in narrow frequency ranges) are opposing constraints. We attempted imaging 32, 64, 128 and 256\,MHz band images and found that below 128\,MHz the images were dominated by deconvolution errors resulting from insufficient $uv$-coverage to image the complex structure of the radio lobes of NGC612. Therefore we decided on 128\,MHz as the best frequency resolution we can achieve with this data. 
This provided a nominal RM resolution of $\approx$76~rad~m$^{-2}$ and a maximum detectable RM of $\approx$370~rad~m$^{-2}$. In this case, bandwidth depolarization caused by averaging over 128\,MHz is negligible considering the mean RM of $\approx$2~rad~m$^{-2}$ in the direction of this source. 

\begin{figure}
\includegraphics[width=\columnwidth,clip=true,trim=0.0cm 1.0cm 0.0cm 2.0cm]{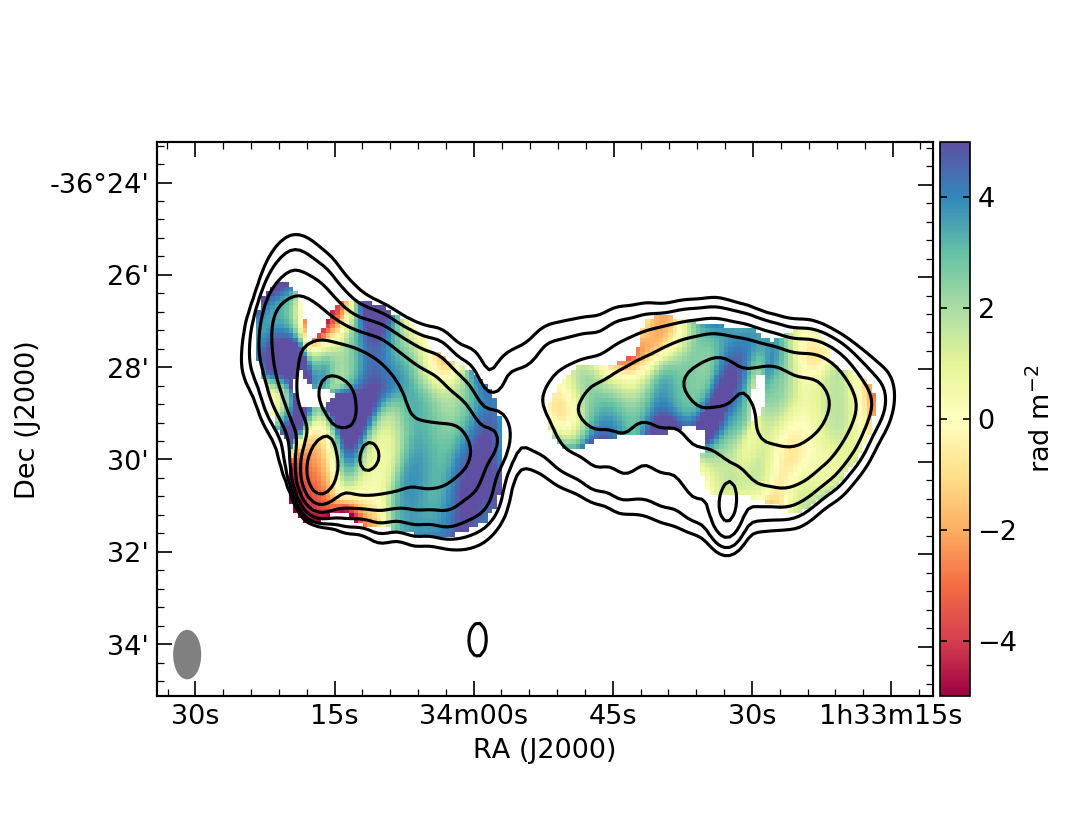}
\includegraphics[width=5.2cm,angle=270]{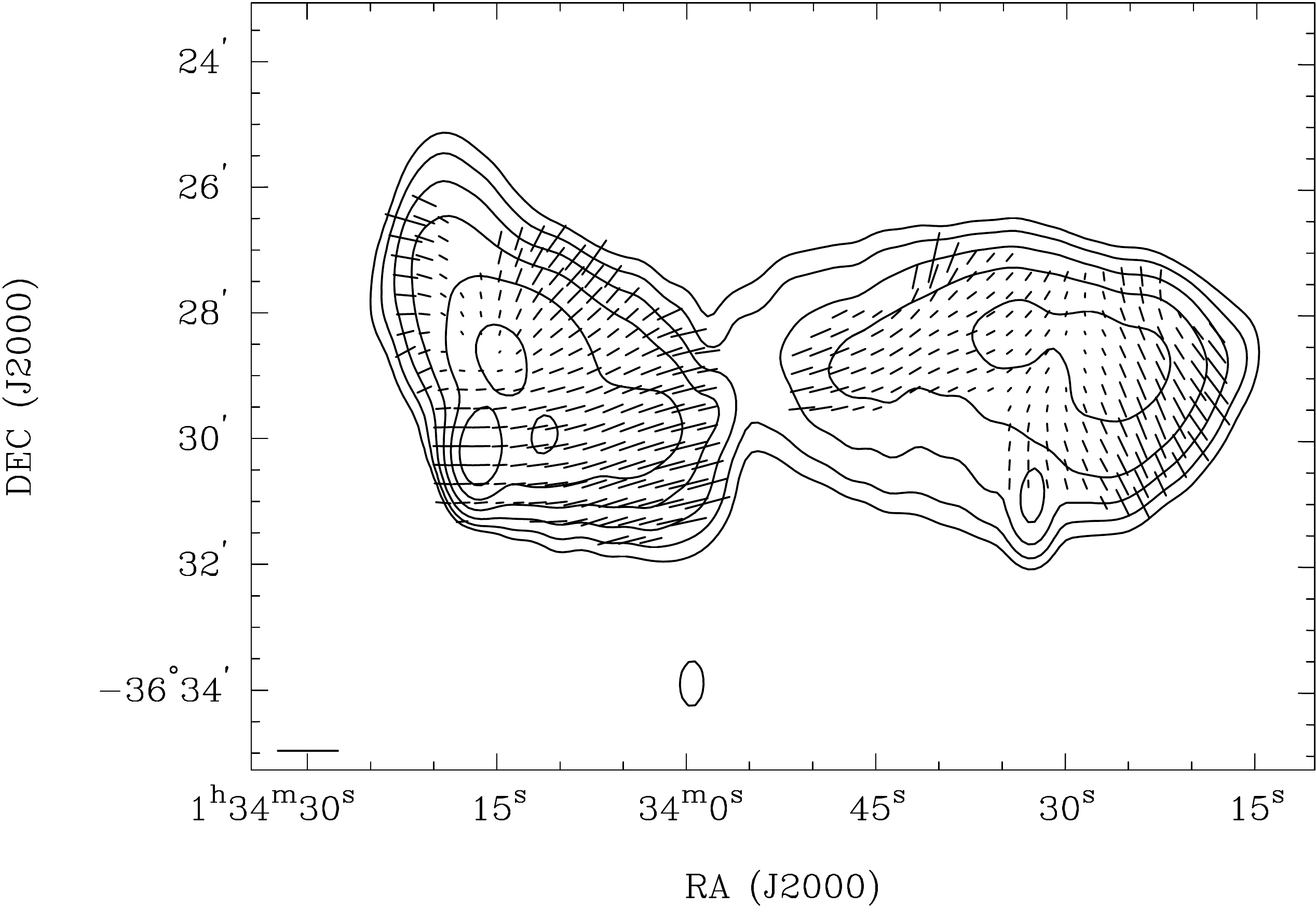}\caption{Top: The RM distribution of NGC\,612 in \rad. The solid black contours show the Stokes $I$ emission 
starting at $4\sigma_{\rm I}$ and increasing by factors of 2.  The synthesised beam is shown in the bottom left corner of the image. 
Bottom: Intrinsic polarization angle ($\Psi_0$) map after correction for Faraday rotation. The vectors represent the orientation 
of $\Psi_0$ with the length corresponding to the relative fractional polarization. A vector 
length corresponding to a fractional polarization of 1 is shown in the bottom left corner of the image. 
A colour version of this figure is available online.}\label{fig:rm}
\end{figure}

\begin{figure*}
\includegraphics[width=15cm]{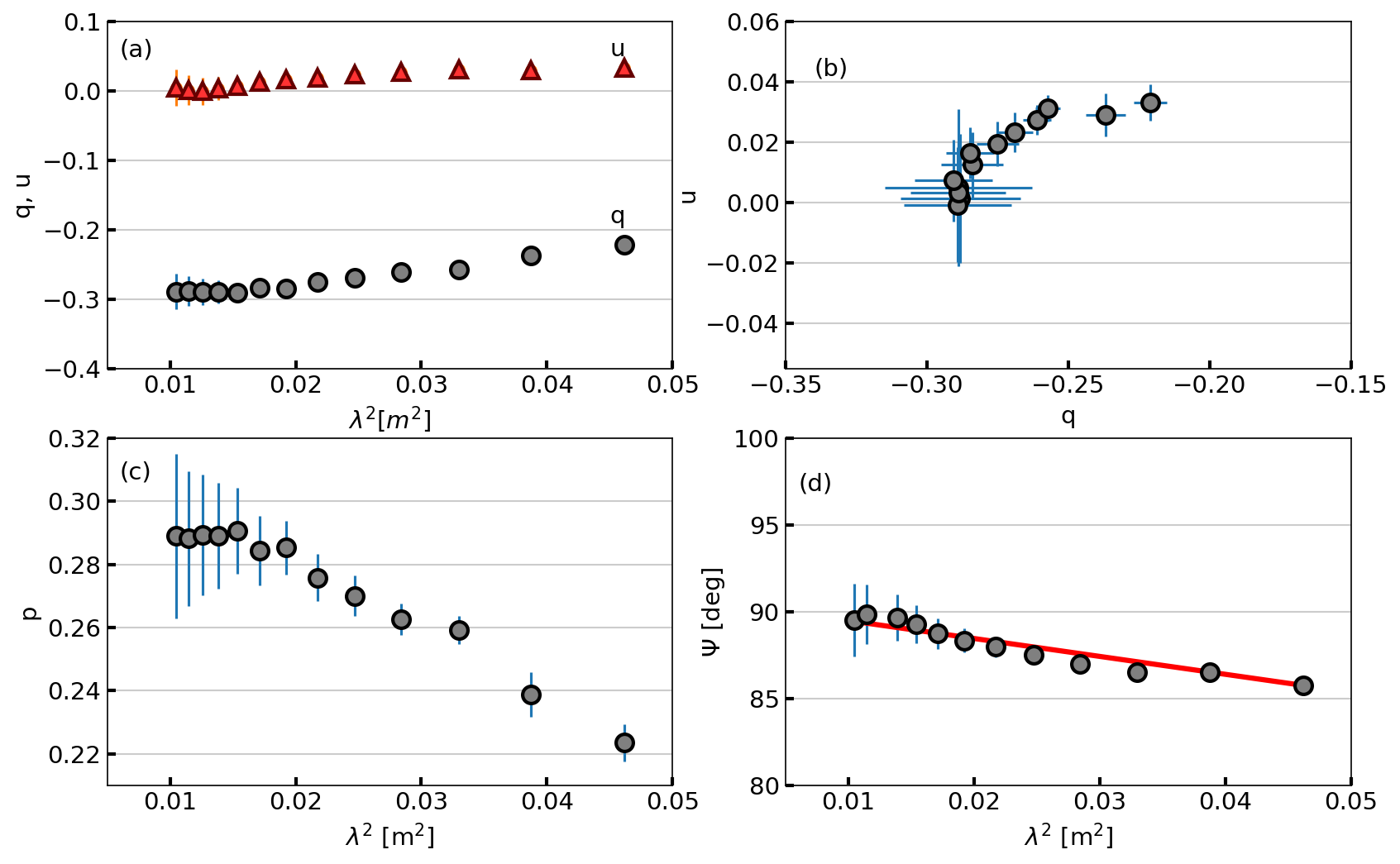}
\caption{The polarization data at the location of the hotspot. (a) $q$ and $u$ as a function of wavelength squared. (b) $q$ versus $u$. (c) Fractional polarization versus wavelength squared. (d) Polarization angle versus wavelength squared, overlaid with a red solid line corresponding to an RM of $-1.8$\rad, as derived from RM synthesis at that location (Figure~\ref{fig:rm}). A color version of this figure is available online.}
\label{fig:hotspot}
\end{figure*}

\begin{figure*}
\includegraphics[scale=0.5]{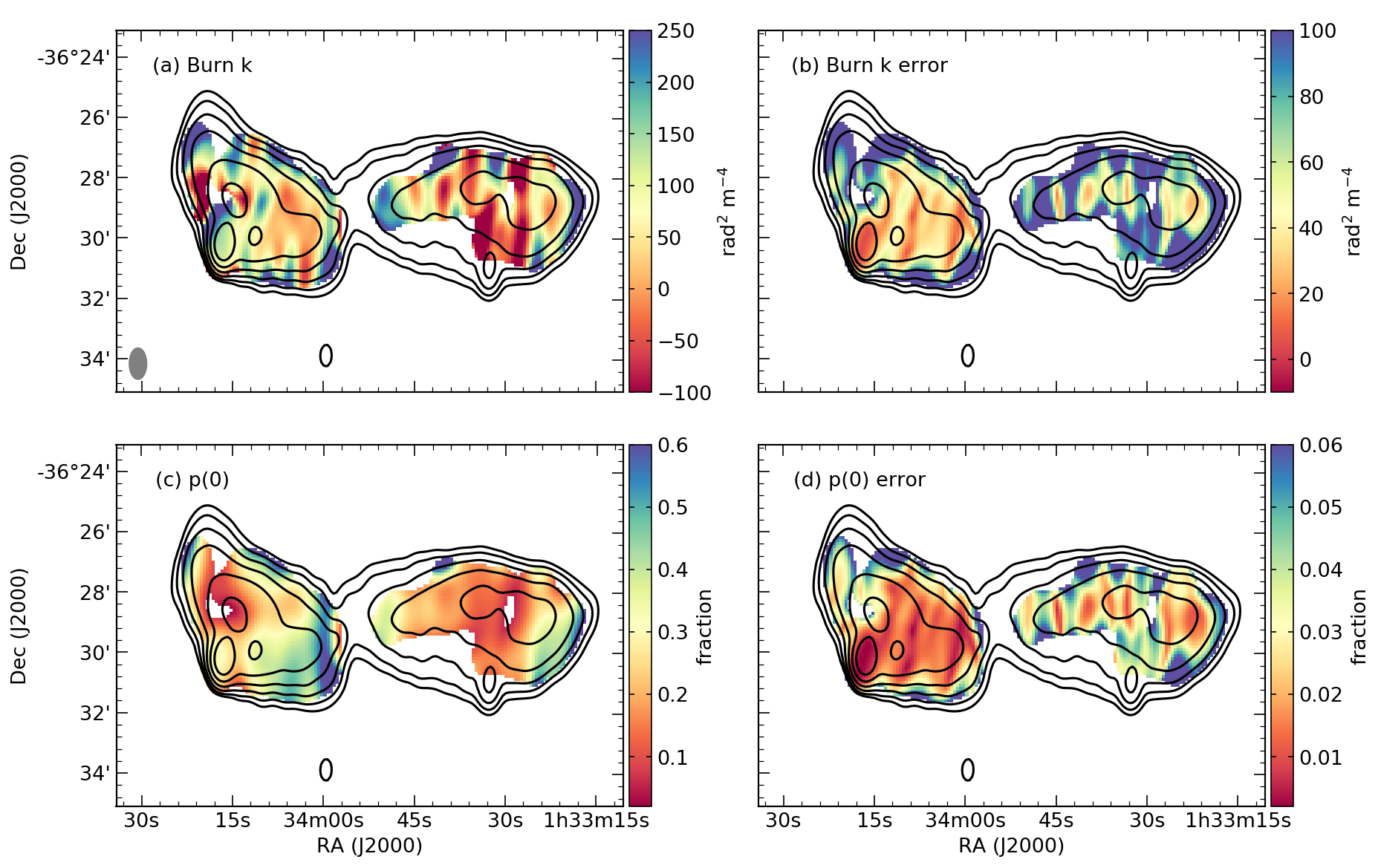}
\caption{Representation of the depolarization of NGC\,612 using the Burn law (Section~\ref{sec:kmap}). (a) The fitted depolarization paramter $k$ from the Burn law. (b) The error in $k$. (c) The intrinsic polarization $p(0)$ from the extrapolation of the Burn law to zero wavelength. (d) The error in $p(0)$. The grey oval in the lower left of (a) represents the beam of the Stokes $I$ emission shown with the black contours.  The contours start at $4\sigma_{\rm I}$ and increase by a factor of 2.  A color version of this figure is available online.}\label{fig:kmap}
\end{figure*}

\begin{figure}
\includegraphics[scale=0.55]{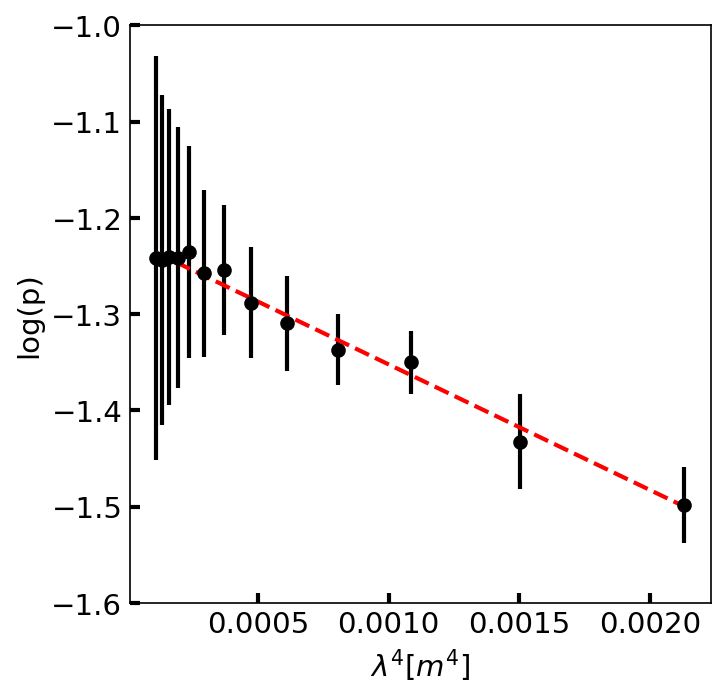}
\caption{Plot of the natural log of the fractional polarization $p$ versus $\lambda^4$. 
The dashed red line represents the best fitting slope of $k=124\pm8$\,rad$^2$~m$^{-4}$. 
A color version of this figure is available online.}\label{fig:kmap-fit}
\end{figure}

\section{Analysis}\label{sec:analysis}
\subsection{Rotation Measure}\label{sec:rm}
Faraday rotation provides information on the magnetic field strength and geometry and ionized particle distributions of magnetised thermal material along the line of sight. Complicated Faraday structures can exist in both compact radio sources \citep[i.e.][]{OSul2012} and from diffuse radio emission found in the lobes of radio sources \citep[i.e.][]{Guidetti2012}. The origin of such RM complexity is still under debate, but it is expected that RM variations caused by the local source environment are a key factor \citep[e.g.][]{osullivan2017, anderson2018}. 
One can potentially use the RM to determine the type of interaction between the radio lobe plasma and the external medium and constrain the magnetic field properties of the intergalactic gas \citep[e.g.][]{Laing2008,guidetti2011}. To determine the distribution of polarization and RM across the lobes, we applied the technique of RM synthesis \citet{Brentjens2005} on the 16 x 128\,MHz images (Section \ref{sec:polarization}). 

The RM distribution, shown in Figure~\ref{fig:rm}, was produced from the fractional $Q$ and $U$ data and pixels were blanked where the peak polarized intensity was less than 8$\sigma_{QU}$ (where $\sigma_{QU}$ was determined from the $Q$ and $U$ RM synthesis spectrum far from the peak). The mean RM of the eastern lobe is $+2.6$~rad~m$^{-2}$ with a standard deviation of 2.6~rad~m$^{-2}$. 
However, the RM is negative at the hotspot and increases in magnitude towards the leading edge (from $-1.8$\rad~to $-5$\rad). 
Figure~\ref{fig:hotspot} shows the behaviour of the polarization data at the hotspot, with the RM of $-1.8\pm0.1$\rad~overlaid on the data in the angle versus wavelength-squared panel. 
For the western lobe, the mean and standard deviation in RM are $+1.7$~rad~m$^{-2}$ and 1.8~rad~m$^{-2}$, respectively. 
The RM increases along the ridge from $\sim$0\rad~in the inner lobe to $\sim$5\rad~at the bend and back down to $\sim$1\rad~towards the end point of the lobe. The mean RM error is 0.3\rad.  

\subsection{Depolarization}\label{sec:kmap}
The high fractional polarization observed across the lobes in the 16~cm band, 
with a mean of $0.23$ and a standard deviation of 0.11, is consistent 
with the small RM variations observed across the majority of the lobes on scales larger than the synthesised beam 
(Section~\ref{sec:rm}). However, there are significant variations in the fractional polarization 
with frequency in particular locations, indicating substantial RM variations on smaller scales 
(e.g.~Figure~\ref{fig:hotspot}). 

To better quantify these variations, we parameterise the depolarization across the radio lobes of NGC\,612 
using the Burn law \citep{Burn1966}, 
which is also commonly known as external Faraday dispersion \citep{Sokoloff1998}. 
We follow a procedure similar to that described in \citet{Laing2008}, where we fit the observed data with the relation 
\begin{equation}
{\rm ln}\,p(\lambda) = {\rm ln}\,p(0)-k\lambda^4 \,\,\, ,
\end{equation}
where $\lambda$ is the wavelength (m) of our observations, $k$ is the depolarization parameter, 
and $p$(0) is the intrinsic polarization. Figure.~\ref{fig:kmap-fit} shows the fit for the hotspot. 
We only use those pixels where $\sigma_p>4$ in all frequencies and include an additional systematic 
error based on a linear interpolation of the widefield instrumental polarization maps (Section~\ref{sec:offpol}). 
We plot the Burn $k$ parameter and the intrinsic polarization maps in Figure~\ref{fig:kmap}, along with their 
corresponding formal error maps.

The observed depolarization differs significantly between the two radio lobes.  
The eastern \FRII\,\,radio lobe has a mean $k$ value of $100$\,rad$^2$~m$^{-4}$ with a standard deviation 
of $124$\,rad$^2$~m$^{-4}$ and a mean error of $70$\,rad$^2$~m$^{-4}$. At the location of the hotspot 
$k=124\pm8$\,rad$^2$~m$^{-4}$ indicating significant depolarization, while $k$ decreases upstream 
towards the host galaxy without much evidence for large depolarization (typical values of $k\sim40$\,rad$^2$~m$^{-4}$ 
with errors of $\sim30$\,rad$^2$~m$^{-4}$). 
The western \FRI\,\,radio lobe shows evidence for a banded depolarization and repolarization structure 
along the radio lobe, however the significance of the variations are low. The mean $k$ value of $70$~rad$^2$~m$^{-4}$ 
is lower than the eastern lobe, but the standard deviation is higher at $160$~rad$^2$~m$^{-4}$. 
The large mean error in $k$ of $100$~rad$^2$~m$^{-4}$ means that these variations are significant at  
about a 90\% confidence level. However, there is also an east-west ripple pattern visible in the total intensity, 
so these variations could be related to imaging artefacts. 
In general, there appears to be more complex polarization behaviour than described by the Burn law, and 
more detailed modelling of the radio lobes \citep[e.g.][]{kaczmarek2018} with higher sensitivity and 
importantly, higher fidelity observations being required to investigate this further. 

At the hotspot, $p(0)=0.295\pm0.003$, and the mean value of $p(0)$ across the eastern lobe is $\sim$0.3 with a standard deviation of 0.15, while the mean error is $\sim$0.03. This is comparable to the $p(0)$ mean and standard deviation of 0.24 and 0.14, respectively, of the western lobe (with a mean error of $\sim$0.04). 
The intrinsic fractional polarization increases to as much as 0.7, close to the theoretical maximum, 
at the lobe edges (albeit with relatively large errors of $\sim$0.1). 
In Figure~\ref{fig:rm} (bottom), we show the intrinsic polarization angle ($\Psi_0$) map, after correction for Faraday rotation, and with 
the polarization vectors scaled by the intrinsic fractional polarization. The $\Psi_0$ distribution implies a circumferential magnetic field 
at the edges of the radio lobes, as is commonly observed \citep[e.g.][]{leahy1986}. 
The southern edge of the eastern lobe is an exception to this, with a magnetic field perpendicular to the total intensity contour lines. 
The field in the central lobe regions is orientated approximately along the ridge axes, with this alignment maintained even through the 
significant bend in the western lobe. 
The field orientation at the hotspot is not exactly perpendicular to the presumed jet direction, 
although observations at higher angular resolution that resolve the hotspot region are needed for further study. 

\begin{figure*}
\includegraphics[width=\textwidth]{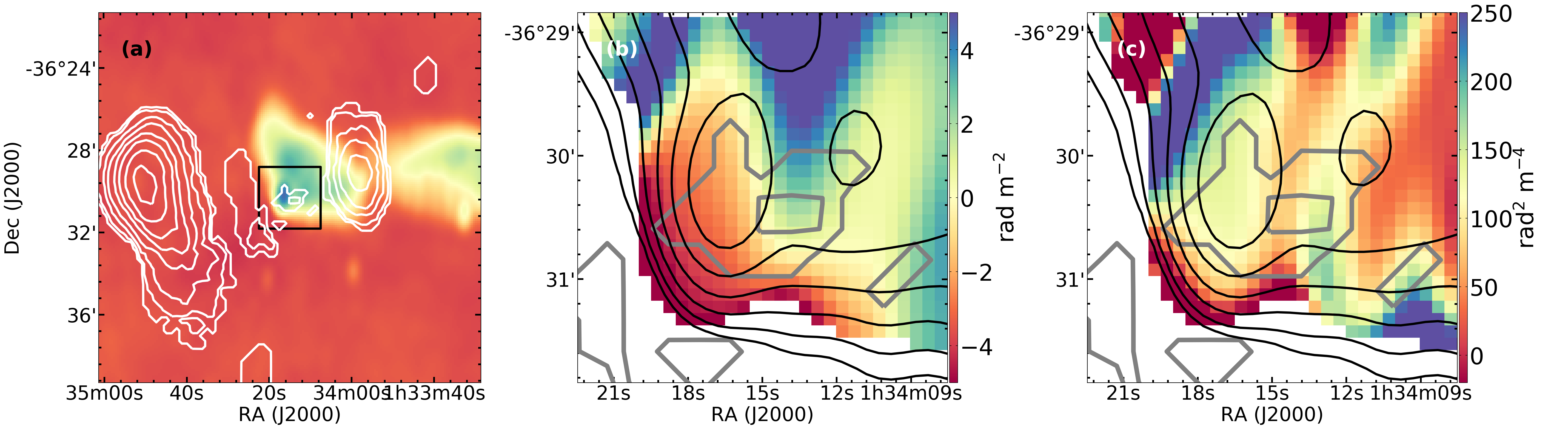}
\caption{The \HI\,\,emission from \citet{Emonts2008} in relation to the radio continuum emission of NGC\,612. (a) Contours of the \HI\,\,emission in and between NGC\,612 and NGC 619 overlaid into the Stokes $I$ continuum emission. 
Contour levels: 0.05, 0.14, 0.24, 0.4, 0.6, 0.9, 1.2, 1.6, 2.0~Jy\,beam$^{-1}$\,$\times$\,km\,s$^{-1}$. The \HI\ emission in the bridge is faint and difficult to accurately image in this total intensity plot, but it show up as a distinct feature in the position-velocity maps of \citet{Emonts2008}. The background color image is the Stokes $I$ continuum image. The black square indicates the zoomed-in region of the (b) and (c) panels. 
(b) The background color image shows the RM map with \HI\,\,emission in grey contours and Stokes $I$ emission in black contours. 
(c) Similar to (b) with the colormap now showing the Burn law $k$ parameter representing Faraday depolarization. 
A color version of this figure is available online.}
\label{fig:HIbridge}
\end{figure*}

\section{Discussion}\label{sec:discussion}
Here we discuss the polarization and Faraday rotation properties of NGC~612 in relation to the details of the interaction between this hybrid morphology radio source and the gas in the circumgalactic environment. 
Large scale mixing of ionized thermal material with the relativistic material in the radio lobe would produce internal Faraday rotation \citep[e.g.][]{Cioffi1980}. This effect has a unique RM and depolarization signature, for sufficient wavelength coverage, that can be detected using RM synthesis as well as through detailed modelling of the variation in both the polarization angle and amplitude with wavelength squared \citep[e.g.][]{anderson2018}. 
We find no clear evidence for internal Faraday rotation in either lobe from the current data, although the western lobe does show a complex depolarization structure that should be investigated further with higher fidelity observations, such as those that will be attained by the Australian Square Kilometre Array Pathfinder \citep[ASKAP;][]{johnston2009}.  

In the case of the circumgalactic gas mixing at the lobe boundary rather than mixing throughout the lobe volume, then the RM structure is also expected to be correlated with the source structure. 
For example, \cite{Bicknell1990} considered a model where any emission from the boundary layer is completely depolarized and the RM structure correlates with the lobe structure without any large variation in the degree of polarization with wavelength. 
The western lobe shows a significant increase in RM magnitude at the location of a distinct bend in the lobe. This suggests that the ambient gas responsible for the observed Faraday rotation is in the local source environment and being influenced by the radio lobe. 
However, an RM structure correlated with the source structure does not directly imply a mixing layer. 
Recently, \citet{Guidetti2012} discovered coherent RM structures across the lobes of several nearby radio galaxies. They postulated that these could be potentially generated by compression along with draping and stretching of the magnetic field in the ambient gas by the expansion of the lobes \citep[e.g.][]{Dursi2008}. 

In the case of the eastern lobe, the strong shock expected at the hotspot should compress the ambient gas, enhancing the RM contribution local to the source, if the external medium is appreciably magnetised \citep[e.g.][]{Carilli1988}. 
We consider the bow-shock scenario as a possible explanation for the RM structure of the hotspot, where the RM magnitude increases across the hotspot towards the leading edge of the lobe. The amount of depolarization at the hotspot is also $\sim$3 times higher than along the ridge of the lobe. The increase in depolarization makes sense if the magnetic field in the compressed external medium is tangled on scales $\ll 50$~kpc.  
The negative sign of the RM at the hotspot is opposite to the positive RM across the majority of the eastern lobe, also consistent with a significant influence of the radio lobe on the external medium at this location. 

Interestingly, NGC\,612 is connected to NGC 619 through a narrow bridge of tidal debris detected in \HI\ \citep{Emonts2008}. The \HI\ bridge connects these two galaxies along the Eastern radio lobe. 
From an in-depth study of both the \HI\ bridge and the regularly rotating \HI\ disk around HGC 612, \citet{Emonts2008} argue that the bridge is the result of an interaction or collision between NGC\,612 and NGC 619 that occurred at least a Gyr ago. Figure~\ref{fig:HIbridge} visualizes the faint \HI\,\,emission from the bridge in relation to the Eastern \FRII\,\,radio lobe and the hotspot. 
It is likely that the majority of the \HI\ emitting material is located behind the radio source. If it were in front of the radio emission, we would have observed the \HI\ gas in the line of sight towards the radio hotspot in absorption rather than emission. 

The presence of an \HI\ bridge indicates that there is likely a larger reservoir of tidal debris between NGC\,612 and NGC\,619. 
If part of this tidal debris was shock ionised by the expanding radio lobe, and in the foreground, then there should be a signature of this ionised gas in the Faraday rotation and depolarization maps. 
Figure~\ref{fig:HIbridge}(b) and (c) shows a zoom-in of the hotspot region with the \HI\ contours overlaid on the RM map and depolarization map, respectively. 
As there is no clear correspondence, we can conclude that any significantly ionised tidal debris is located behind the radio lobe, consistent with the eastern jet and radio lobe orientated either close to the plane of the sky or towards the line of sight. While we cannot rule out an interaction between the radio lobe and the tidal debris, the RM properties discussed in this paper, combined with the fact that the \HI\ in the bridge has a very narrow velocity dispersion \citep{Emonts2008}, provide no clear evidence for this.
Deeper \HI\ studies in combination with X-ray and H$\alpha$ observations would help further illuminate the complex gas dynamics of this system. 

\section{Summary}
We have presented an analysis of the interaction between the lobes of the hybrid morphology radio galaxy, NGC\,612, and the circumgalactic gas, through a broadband radio polarization and Faraday rotation study, using the Australian Telescope Compact Array at 1 to 3\,GHz. Previous \HI\ observations revealed the presence of a bridge of tidal debris connecting NGC\,612 to a nearby galaxy. The tidal bridge is coincident on the sky with the eastern FRII lobe, with no \HI\ detected towards the western FRI lobe. 

We examined the linear polarization, Faraday rotation and depolarization distributions, finding that they are consistent with external Faraday rotation caused by a turbulent magnetic field in the intragroup medium. 
There is evidence for a correlation between the Faraday structure and the radio morphology, both at the location of the hotspot in the eastern lobe and at a bend in the western lobe. 
In particular, the enhancement in RM and depolarisation at the hotspot can be plausibly explained by the projection of part of the bow-shock region onto the lobe, where the ambient gas has been appreciably compressed.
Although several aspects of the RM and depolarization structure suggest an influence of the radio galaxy on the ambient magnetoionic environment, there is no evidence for either large scale mixing of the eastern radio lobe with the gas in the \HI\ bridge, or large amounts of ionised debris in the foreground of the radio lobe. 

After correction for the effect of Faraday rotation, we find a mainly circumferential intrinsic magnetic field structure of the lobes (the southern edge of the eastern lobe is a notable exception to this). This is a commonly observed feature of double-lobed radio sources, and is consistent with an initially tangled magnetic field that is stretched by the shearing motion along the direction of the backflow, in addition to outward compression due to the lobe expansion. 

In general, broadband radio polarization observations with higher image fidelity and sensitivity are required to better determine the nature of the relationship between the Faraday rotating gas and the radio lobe. Complementary H$\alpha$ and X-ray observations would also provide a better understanding of the multi-phase intragroup medium surrounding this hybrid morphology radio galaxy. 
In general, this work demonstrates some aspects of how broadband radio polarization data can provide a means to examine the detailed role of AGN feedback in environments other than galaxy clusters at low redshift. Further studies of NGC\,612 and other galaxies in similar environments can provide new insights into radio galaxy evolution and feedback across a wide range of redshifts. 

\section*{Acknowledgments}
We would like to thank R.~Ekers for useful comments and discussion. Parts of this research were conducted by the Australian Research Council Centre of Excellence for All-sky Astrophysics (CAASTRO), through project number CE110001020. The Australia Telescope Compact Array is part of the Australia Telescope, which is funded by the Commonwealth of Australia for operation as a National Facility managed by CSIRO. The National Radio Astronomy Observatory is a facility of the National Science Foundation operated under cooperative agreement by Associated Universities, Inc. The figures in this work made use of Astropy, a community-developed core Python package for Astronomy \citep{astropy2013}. 
SPO acknowledges financial support from the Deutsche Forschungsgemeinschaft (DFG) under grant BR2026/23.

\bibliographystyle{mnras}

\label{lastpage}

\end{document}